\documentclass{myjfm}
\usepackage{natbib}
\usepackage{graphicx}
\usepackage{dcolumn}
\usepackage{amssymb,amsmath}
\usepackage{epsfig}
\usepackage[dvips]{color}
\usepackage{lscape}
\usepackage{longtable}
\usepackage{rotating}

\usepackage{hyperref}
\hypersetup{
breaklinks=true,
hyperfootnotes= true,
pagecolor=white,
colorlinks=true,
linkcolor= black,
citecolor= black,
urlcolor=blue
}
\def\href#1#2{{#2}}

\newcommand\euro{{\sffamily C%
    \makebox[0pt][l]{\kern-.70em\mbox{--}}%
    \makebox[0pt][l]{\kern-.68em\raisebox{.25ex}{--}}}~}

\def\be{\begin{equation}}
\def\ee{\end{equation}}

\def\x{{\mbox{\boldmath$x$}}}
\def\u{{\mbox{\boldmath$u$}}}

\def\v{{\mbox{\boldmath$v$}}}

\def\bx{{\mathbf{x}}}
\def\cA{{\mathcal{A}}}
\def\cB{{\mathcal{B}}}

\def\begineq{\begin{equation}}
\def\endeq{\end{equation}}

\title{Dimensionality and morphology of particle and bubble 
clusters in turbulent flow}
\author[Calzavarini, Kerscher, Lohse, and Toschi]{
Enrico Calzavarini$^{1,4}$, Martin Kerscher$^2$, Detlef Lohse$^{1,4}$, 
and Federico Toschi$^{3,4}$}
\affiliation{
$^1$Physics of Fluids Group, 
Department of Science and Technology, 
J.M. Burgers Center for Fluid Dynamics, and Impact-Institute, \\
University of Twente, P.O Box 217, 7500 AE Enschede, The
Netherlands,\\
$^2$Mathematisches Institut,
Ludwig--Maximilians--Universit\"{a}t, Theresienstrasse 39, D--80333
M\"{u}nchen, Germany,\\
$^3$IAC-CNR, Istituto per le Applicazioni
del Calcolo, Viale del Policlinico 137, I-00161 Roma, Italy and INFN,
via Saragat 1, I-44100 Ferrara, Italy\\
$^4$International Collaboration for Turbulence Research
}

\date{\today}

\begin{document}

\maketitle

\begin{abstract}
We conduct numerical experiments to investigate the spatial clustering
of particles and bubbles in simulations of homogeneous and isotropic
turbulence. Varying the Stokes parameter and the densities, striking
differences in the clustering of the particles can be observed.
To quantify these visual findings we use the Kaplan--Yorke
dimension. This local scaling analysis shows a dimension of
approximately 1.4 for the light bubble distribution, whereas the
distribution of very heavy particles shows a dimension of
approximately 2.4.  However, clearly separate parameter combinations
yield the same dimensions.
To overcome this degeneracy and to further develop the understanding of
 clustering, we perform a morphological (geometrical and
topological) analysis of the particle distribution.  For such an
analysis, Minkowski functionals have been successfully employed in
cosmology, in order to quantify the global geometry and topology of
the large-scale distribution of galaxies.  In the context of dispersed
multiphase flow, these Minkowski functionals -- being morphological
order parameters -- allow us to discern the filamentary structure of
the light particle distribution from the wall-like distribution of
heavy particles around empty interconnected tunnels.
\end{abstract}

\section{Introduction}
Even in homogeneous and isotropic turbulence particles, drops, and bubbles
(all from now on called ``particles'') do not distribute
homogeneously, but {\em cluster}, see
 \cite{cro96} for a classical review article.
The clustering
has strong bearing on so diverse issues such as aerosols and cloud
formation (\cite{fal02,cel05,vai02}), plankton distribution in the deep
ocean (\cite{mal06}), sedimentation and CO$_2$ deposition in water
(\cite{ups06}). Considerable advances in particle tracking velocimetry
(\cite{lap01,hoy05,bou06,bew06,ayy06}) and in numerics
(\cite{elg92,elg93,wan93,boi98a,dru01a,mar02,maz03a,maz03b,bif05,chu05,ber06b,bec06,bec06b}) now allow for the acquisition of huge
data sets of particle positions and velocities in turbulence.

In our numerical experiments the fluid flow is simulated by the
incompressible Navier-Stokes equation on a $128^3$, a $512^3$, and a
$2048^3$ grid with periodic boundary conditions and a large scale
forcing, achieving  Taylor-Reynolds numbers of $Re_\lambda=75$, $180$
and $400$, respectively.  
One-way coupled point-particles are included which experience inertia forces, added
mass forces, and drag, i.e., the particle acceleration is given by
(\cite{max83,gat83}) 
\be {d \v\over dt} = \beta {D\over Dt } \u (\x (t),t) - {1\over \tau_p} (\v - \u(\x (t),t)),
\label{p-acc}
\ee 
where $\v = d\x/dt$ is the particle velocity and $\u(\x (t),t)$ the velocity field.  
Equation (\ref{p-acc}) holds in the limit of small ($\ll 1$) particle Reynolds number and for particles whose size is 
small as compared to the Kolmogorov length scale $\eta$ -- for finite
size particles the results will naturally differ.  
Lift, buoyancy, two-way-coupling and particle-particle interactions are ignored.  
Next to the Reynolds number the control parameters are the particle radius $a$, the density
of the fluid $\rho_f$ and of the particle $\rho_p$.  The dimensionless
numbers used to characterize the particle in the turbulent flow are
the density ratio $\beta=3\rho_f/(\rho_f+2\rho_p)$ and the Stokes
number $St=\tau_p/\tau_\eta=a^2/(3\beta \eta^2)$, where
$\tau_p=a^2/(3\beta \nu)$ is the particle time scale, $\nu$ the
viscosity, and $\tau_\eta$ the Kolmogorov  time
scale.  
In our numerical study we treat $\beta$ and $St$ as independent parameters.
The case $\beta=0$ (and finite $St$) corresponds to very heavy
particles and $\beta = 3$ to very light particles, e.g.\ (coated)
bubbles (i.e., with no-slip boundary conditions at the interface).
$\beta = 1$ means neutral tracers and if in addition $St=0$ we have
fluid particles.

\begin{figure}
\begin{center}
(a) \hspace{5cm} (b) \hspace{5cm} (c)  
\includegraphics[width=1.0\textwidth]{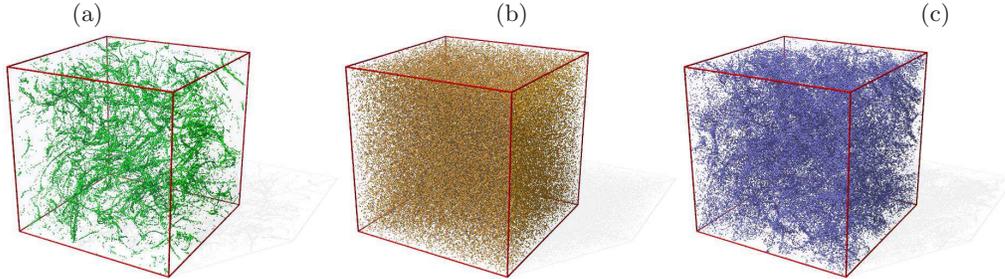}
\caption[]{(color) Snapshots of the particle distribution in the turbulent 
flow field for $St=0.6$ for 
(a) $\beta = 3$ (bubbles),
(b) $\beta = 1$ (tracers), and
(c) $\beta = 0$ (heavy particles), all for $Re_\lambda = 75$.
Corresponding videos are shown in the
accompanying material (or can be downloaded from
\url{http://cfd.cineca.it/cfd/gallery/movies/particles-and-bubbles}).}
\label{snapshots}
\end{center}
\end{figure}

In figure~\ref{snapshots} snapshots of the resulting particle
distributions from a simulation with $Re_\lambda = 75$, $St=0.6$ and
the total number of particles, $N=10^5$ are shown.  For $\beta=0$
(heavy particles) and $\beta=3$ (light bubbles) we observe clustering
-- but of different type. No clustering is observed for the neutral
tracers with $\beta= 1$.

\begin{figure}
\begin{center}
\includegraphics[scale=0.75]{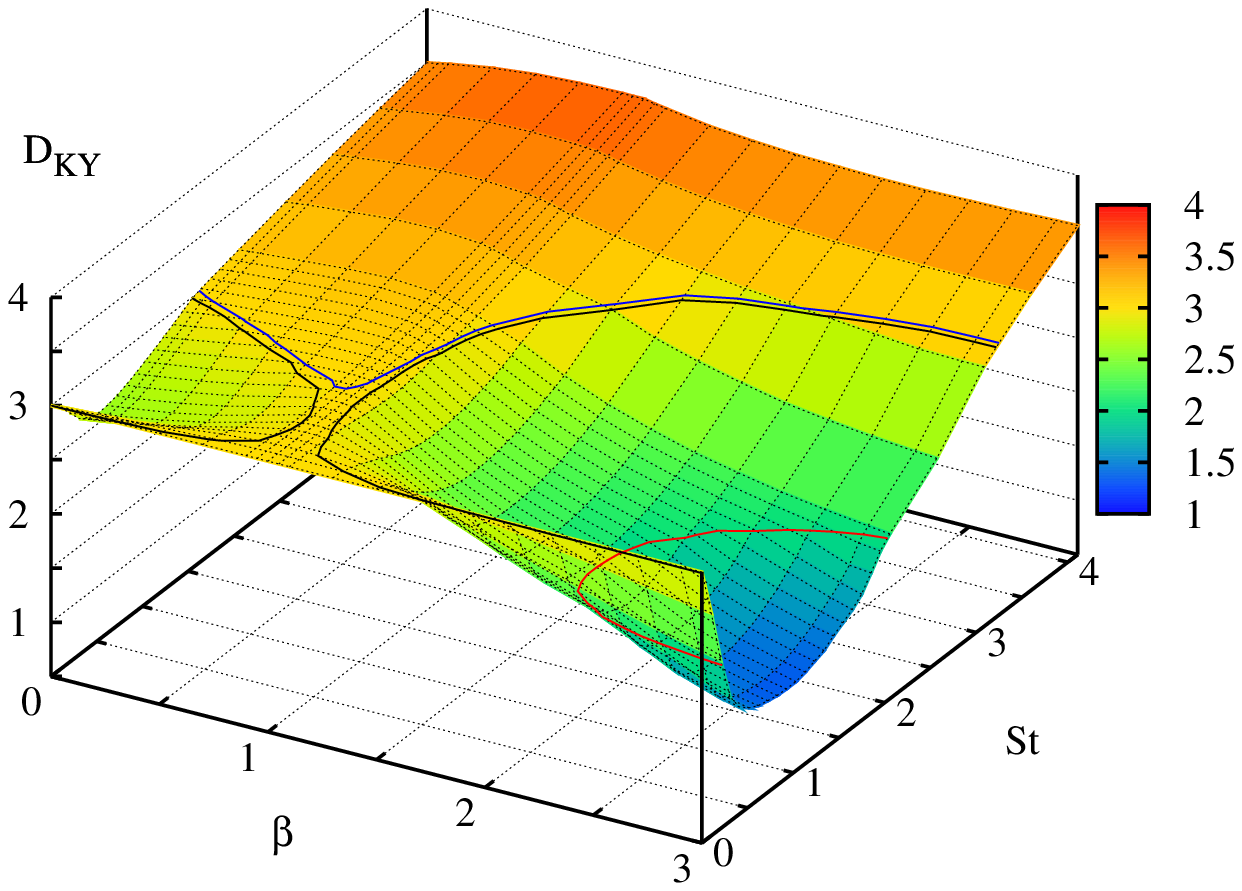}
\includegraphics[scale=0.75]{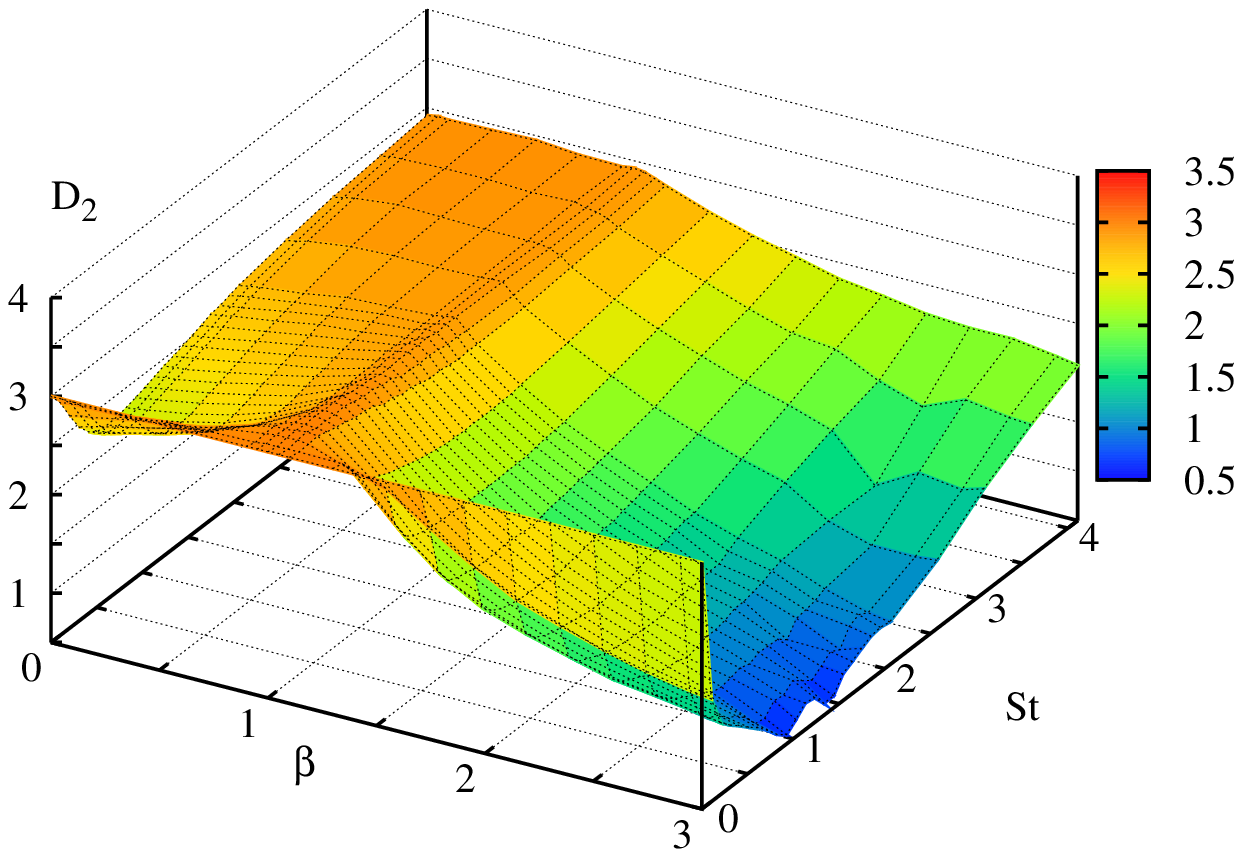}
\caption[]{(color)
The Kaplan-Yorke dimension $D_{KY}$ and
the correlation dimension $D_2$ (bottom) of the particle distribution as
function of $St$ and $\beta$. 
The black line in $D_{KY}$ marks the value $D_{KY}(\beta , St)=3$, the red line the value 2.
For the $D_{KY}$ plot $10^6$ particles were integrated along the tangent 
space and for the $D_2$ plot $5 \cdot 10^7$ particles
were followed.
In both cases the averaging time was tens of large eddy turnovers.
The particles were grouped in about
500 different types characterized by their $(\beta, St)$-values. $Re_\lambda
= 75$.  
}
\label{dky}
\end{center}
\end{figure}

\begin{figure}
\begin{center}
\includegraphics[scale=0.8]{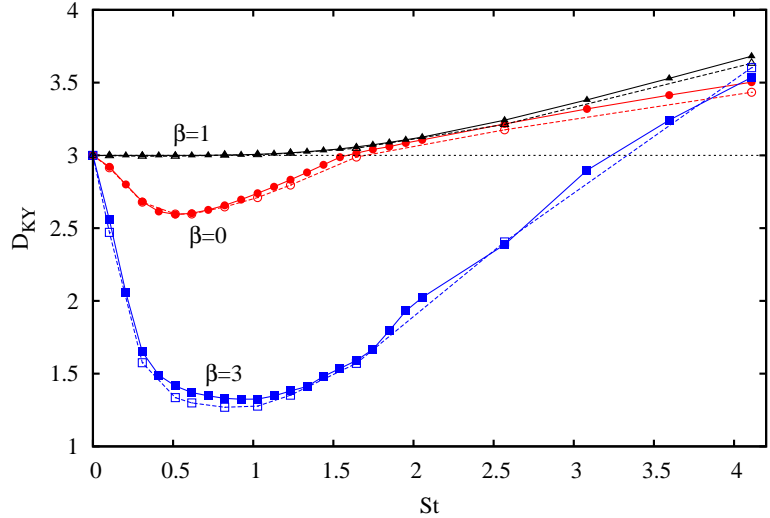}
\caption[]{The Kaplan-Yorke dimension $D_{KY}$ vs.\ $St$ for three $\beta$ values: $\beta=0\ (\circ),1\ (\bigtriangleup), 3\ (\square)$. Results at two different $Re_{\lambda}$ are reported: $Re_{\lambda}=75$ (filled symbols), $Re_{\lambda}=180$ (open symbols).}
\label{dkycut}
\end{center}
\end{figure}

\begin{figure}
\begin{center}
(a) \hspace{4cm} (b) \hspace{4cm} (c)
\includegraphics[width=1.0\textwidth]{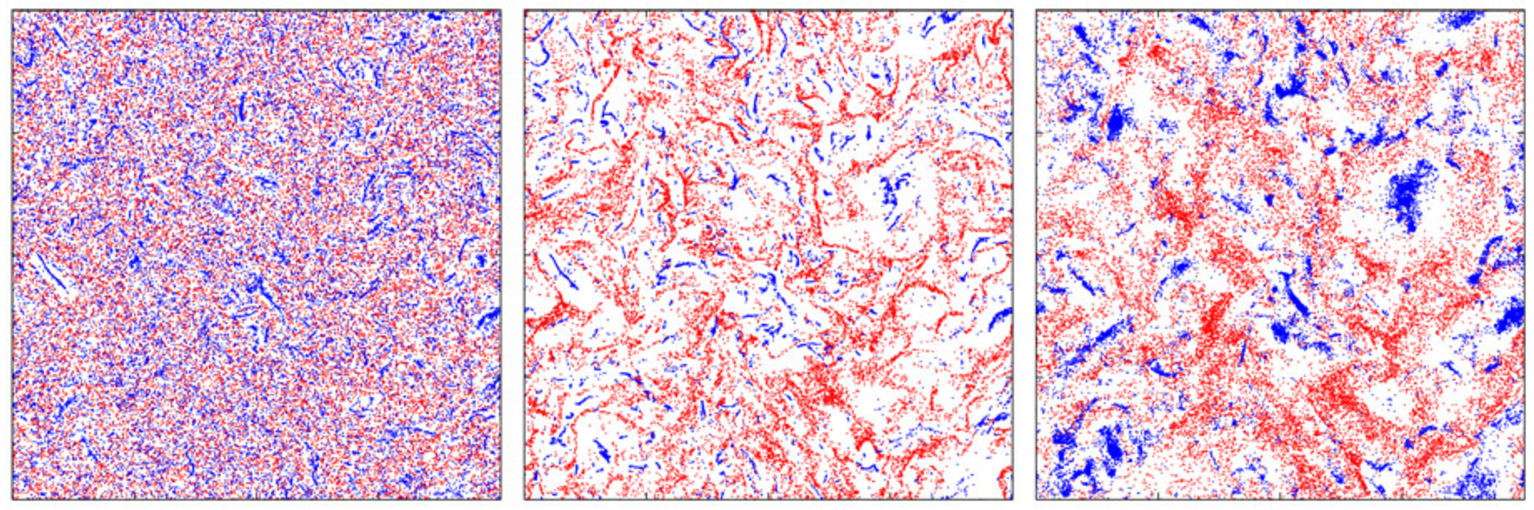}
\caption[]{(color)
Projection of the particles in a slice of $19\eta$ thickness and side
$640\eta \times 640\eta$ with $Re_{\lambda} = 180$,
for
(a) $St = 0.1$,
(b) $St = 0.6$, and 
(c) $St = 4$.
Heavy particles ($\beta = 0$) are shown in red, bubbles ($\beta = 3$) in blue.
To illustrate the different clustering, particles and bubbles in the
same simulation are plotted on top of each other -- note that they
do not mutually interact, i.e., collisions and feedback on the fluid are neglected.
}
\label{projection}
\end{center}
\end{figure}

\section{Dimensionalities of particle and bubble clusters}
How to quantify and characterize the clustering?
One way borrowed from dynamical system theory is to calculate the
Kaplan-Yorke dimension $D_{KY}$ in the six-dimensional space spanned
by the particle positions and their velocities, see
\cite{bec03,bec05}. The Kaplan-Yorke dimension follows from the
Lyapunov exponents (\cite{er85}), $\lambda_i$ ($i\!=\!1,\!\dots,6$), and quantifies how contracting the dynamical system is. It is defined as $D_{\mathrm{KY}}\! =\! K \!+\!  \sum_{i=1}^{K} \lambda_i /
|\lambda_{K\!+\!1}| $, $K$ being the largest integer such that
$\sum_{i\!=\!1}^{K} \lambda_i \geq 0 $. Details on the method can be found in \cite{bec05},
where  the case $\beta=0$ was studied.

In fig.~\ref{dky}, upper,
the  full landscape of $D_{KY}$ as function of $\beta$ and $St$
is shown 
for $Re_\lambda = 75$; 
in fig.\ \ref{dkycut} cuts through this landscape for fixed
$\beta = 0$, 1, and 3 are shown for both  $Re_\lambda = 75$
and $Re_\lambda = 180$, revealing at most a minute Reynolds number
dependence of $D_{KY}$. 

We now discuss the dependence $D_{KY}(\beta, St)$:
Point-particles ($St=0$, $\beta
=1$) do not experience any drag ($\v = \u$) and accordingly $D_{KY} =
3$.  For fixed $\beta$ the contraction is strongest for a Stokes
parameter $St$ in the range between 0.5 and 1.5, i.e., when the
characteristic time scale $\tau_p$ of the particle roughly agrees with
the Kolmogorov time scale $\tau_\eta$. For smaller $St$ the particles
follow the small scale turbulent fluctuations and the clustering
decreases and $D_{KY}$ increases. For large $St$ the particles are so
big that they average out the small scale turbulent fluctuations (see
figure \ref{projection}); correspondingly, the clustering decreases
and $D_{KY}$ increases. The dynamical evolution of the 
heavy particles ($\beta = 0$) shows the
strongest contraction for $St\approx0.5$ with $D_{KY}=2.6$.
Also that of the light
particles ($\beta = 3$) has the strongest contraction for
$St\approx0.5$. Now even  $D_{KY}=1.4$ is achieved.
Figures \ref{dky} and in  \ref{dkycut} reveal another feature
of $D_{KY}$ which is worth being mentioned:
$D_{KY}(St, \beta =1)$ 
slightly increase with increasing $St$, 
reflecting the fact that 
large particle, even if neutrally buoyant, do not exactly follow the
flow, see the discussion in \cite{bar00}. 

An analysis of the spatial distribution performed with the correlations dimension $D_2$
(\cite{gra84}) yields very similar results as that for $D_{KY}$, 
see fig.\ \ref{dky}, lower (see also \cite{bec07} for further details on this observable).  The $D_2$ is computed by fitting the separation probability
$P_2(r)$, which is the probability that the distance between two particles is less than $r$.
It is assumed that $P_2(r) \sim r^{D_2}$ as $r\to 0$.  
As the numerical convergence is more difficult to achieve for
$D_2$, the curve $D_2(\beta, St)$ is slightly less smooth as compared to $D_{KY}(\beta, St)$. 
The finiteness of the particle samples, here roughly 20 snapshots of $10^5$ particles, leads to noisy measures especially for the highly clustered cases ($\beta \sim 3$): Point-particles may reach vanishingly small distances which makes
it difficult to perform a proper histogram binning, and thus
 to compute statistically stable $P_2(r)$.

Due to its local character a metric measurement with the Kaplan--Yorke
or correlation dimension cannot supply us with global morphological
information. Indeed, as seen from figure \ref{dky}, clusters of
heavy and light
particles can have the very same $D_{KY}$, though they look
very different. 
Indeed, 
the striking morphological differences of the particle and bubble
distribution within the same turbulent velocity field are illustrated
in figure \ref{projection}.


\section{Morphology of particle and bubble clusters}
With global geometrical and topological order parameters we are able
to distinguish the clustering on spatially extended, eventually
interconnected, sheets from clouds or filamentary clustering.
Consider the union set $\cA_r=\bigcup_{i=0}^N\cB_r(\bx_i)$ of balls of
radius $r$ around the $N$ particles at positions $\x_i$, $i=1,2,\dots
N$, thereby creating connections between neighboring balls.  The
global morphology of the union set of these balls changes with radius
$r$, which is employed as a diagnostic parameter.
It seems sensible to request that global geometrical and topological
measures of e.g.\ $\cA_r$ are additive, invariant under rotations and
translations, and satisfy a certain continuity requirement. With
these prerequisites  \cite{had57} proved that in three
dimensions the four Minkowski functionals $V_{\mu}(r)$, $\mu =
0,1,2,3$, give a complete morphological characterization of the body
$\cA_r$.
The Minkowski functional $V_0(r)$ simply is the volume of $\cA_r$,
$V_1(r)$ is a sixth of its surface area, $V_2(r)$ is its mean
curvature divided by $3\pi$, and $V_3(r)$ is its Euler
characteristics.
%
Volume and surface area are well known quantities.  The integral mean
curvature and the Euler characteristic are defined as surface
integrals over the mean and the Gaussian curvature respectively.  This
definition is only applicable for bodies with smooth boundaries.  In
our case we have additional contribution from the intersection lines
and intersection points of the spheres.  \cite{mec94} discuss the
definitions of the integral mean curvature and the Euler
characteristic for unions of convex bodies.  The Euler characteristic
as a topological invariant allows for several other definitions, like
$\chi = \#(\text{isolated bodies}) - \#(\text{tunnels}) +
\#(\text{completely enclosed cavities})$.
Minkowski functionals have been developed for the morphological
characterization of the large scale distribution of galaxies by
\cite{mec94} and have successfully been used in cosmology by
\cite{ker97,ker00,ker01}, porous and disordered media by \cite{arn02},
dewetting phenomena by \cite{her98}, statistical physics in general
(\cite{mec00}) and have been recently employed to study magnetic
structure in small scale dynamos (\cite{wil07}).

There is an interesting relation to fractal dimensions. The scaling
behaviour of the volume density for $r\rightarrow0$ gives the
Minkowski-Bouligand dimension of a Fractal which equals the
box-counting dimension, which is again an estimate of the Hausdorff
dimension (see e.g. \cite{fal90}).  However due to discreteness
effects the scaling behaviour of the volume only yields a noisy
estimate of the dimension. The $D_2$ and $D_{KY}$, as used in our
case, are much more reliable measures for the dimension.
Since the Minkowski functionals are dimensional quantities, one is
able to define other scaling dimensions beyond the volume.
This has been formalized and investigated for special models by
\cite{mec00}. It remains an open question how to apply this to points
decorated with spheres.

\begin{figure}
\begin{center}
\includegraphics[scale=0.5]{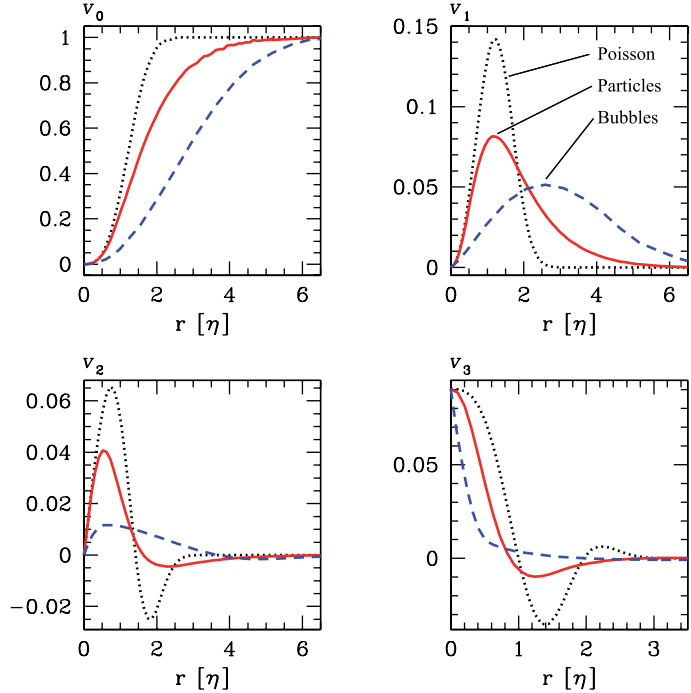}
\caption{ (color) Volume densities of the Minkowski functionals
 $v_{\mu}(r)$, $\mu = 0,1,2,3$, for passive tracers ($\beta = 1$,
 black dotted, corresponding to those of a Poisson distribution
 \cite{mec91}), heavy particles ($\beta = 0$, red solid), and bubbles
 ($\beta = 3$, blue dashed). In all cases $St=0.6$ and $Re_\lambda =
 78$.  Distances are in Kolmogorov units $(\eta)$.  To estimated the
 errors in the Minkowski functionals we looked at several realizations
 of a Poisson process. The one sigma error bars are smaller than the
 linethickness of the shown curves which illustrates the robustness of
 the Minkowski functionals.
-- The code used for the calculations of the Minkowski functionals is
an updated version of the code developed by \cite{ker97}, based on the
methods outlined in \cite{mec94}.  The code is made available to the
general public via
\url{http://www.mathematik.uni-muenchen.de/~kerscher/software/} .}
\label{min}
\end{center}
\end{figure}

In figure \ref{min} we show the volume densities of the four Minkowski
functionals $v_\mu(r)= V_\mu (r)/L^3$, 
$\mu = 0,1,2,3$, determined from the particle
distributions with $St=0.6$ for the three cases with $\beta=3$
(``bubbles''), $\beta=0$ (heavy particles), and $\beta=1$ (neutral
tracers). 
For the neutral tracers ($\beta=1$) the functionals coincide with the
analytically known values (\cite{mec91}) for randomly distributed objects.
As the radius increases, the volume is filled until reaching complete
coverage where the volume density $v_0(r)$, i.e.\ the filling factor,
reaches unity.  This increase is considerably delayed for heavy
particles and even more for bubbles, which is a characteristic feature
of a clustering distribution produced from the empty space in between
the clusters. 

The density of the surface area, measured by $v_1(r)$, increases with
the radius $r$.  As more and more balls overlap the growth of $v_1(r)$
slows down and the surface area reaches a maximum. For large radii the
balls fill up the volume and no free surface area is left.  For both
the bubbles and the heavy particles the maximum of $v_1(r)$ is smaller
compared to the Poisson case.  For intermediate and large radii we
observe the skewed shape of $v_1(r)$ with a significant excess of
surface area on large scales compared to Poisson.  The particles
cluster on clumpy, filamentary and sheet like structures and the
surface area of the balls is growing into the empty space in between.
Especially for the bubbles ($\beta=3$) the maximum of $v_1(r)$ is
attained for considerably larger $r$ compared to the distributions
with $\beta=0,1$, suggesting mainly separated filamentary shaped
clusters.

The density of the integral mean curvature $v_2(r)$ allows us to
differentiate convex from concave situations.  For small radii the
balls are growing outward. The main contributions to the integral mean
curvature is positive, coming from the convex parts. Increasing the
radius further we observe a maximum for all three cases, but as with
the surface area, the amplitude of the maximum is reduced for heavy
particles and especially for the bubbles.
For tracer particles ($\beta=1$) the empty holes start to fill up and
the structure is growing into the cavities.  Now the main contribution
to the integral mean curvature is negative, stemming from the holes
and tunnels through the structure. This concaveness is less pronounced
for the heavy particle case ($\beta=0$). Typically interconnected
networks of tunnels show such a reduced negative contribution.  For
bubbles ($\beta=3$) $\cA_r$ is hardly concave, i.e., no holes and no
tunnels develop, just as expected from isolated (filamentary)
clusters.  For large radii the balls fill up the volume, no free surface
and hence no curvature is left.

The topology undergoes a number of changes which we measure with the
Euler characteristic. For small radii $r\approx0$ the balls remain
separated and the volume density of the Euler characteristic
$v_3(r\approx0)$ equals the number density of the particles. As the
radius increases, balls join and the Euler characteristic decreases.
Both for heavy particles ($\beta=0$) and especially for bubbles
($\beta=3$) the decrease of $v_3(r)$ with increasing $r$ is more
dramatic, due to the clustering.
When further increasing the radius $r$, more and more tunnels start to form
resulting in a negative $v_3(r)$. This is observed for neutral particles
($\beta=1$) and less pronounced for the heavy particles
($\beta=0$). No tunnels seem to form in the bubble distribution
($\beta=3$).
For neutral particles this behavior reaches a turning point when these
tunnels are blocked to form closed cavities and a second positive
maximum of $v_3(r)$ can be seen.  Neither bubbles nor heavy particles
show a significant positive $v_3(r)$.

To develop a physical picture we first look at the heavy
particles. One expects that the particles are dragged out of the
turbulent vortices gathering on structures around the vortices. A
dimension $D_{KY}\approx2.4$ is indicating ``fat'' locally two
dimensional structures.
From the negative $v_2(r)$ for large $r$ we conclude that these
2d-structures are mainly concave, i.e.\ growing inward.  Hence these
2d-structures are not small 2d-patches, but enclose the vortices.
The particles around the vortices form the enclosure of tunnels but
they do not fully enclose the vortices 
(they do not form the ``casing of a
sausage''). Completely anclosed cavities would leed to a $v_3(r)$ larger
zero for large $r$, which we do not observe.  There have to be regions
connecting the vortices devoid of any particles. In other words there
exists an empty interconnected network of tunnels surrounded by
particles.

For the bubble distribution one expects that the bubbles are sucked into
the vortices. The dimensional analysis with $D_{KY}\approx1.4$ is now
suggesting ``fat'' locally one dimensional structures.
For the bubble distribution $v_2(r)$ is almost everywhere positive,
hence the 1d-structures remain convex for all radii. Similar $v_2(r)$
is positiv on all scales.  Hence, neither tunnels devoid of bubbles
nor empty cavities enclosed by bubbles form.  The 1d-structures built
from the bubbles are separated filamentary clusters.  This also shows
that the bubbles do not form a percolating structure throughout the
simulation.


\begin{figure} 
\begin{center}
  \includegraphics[scale=0.5]{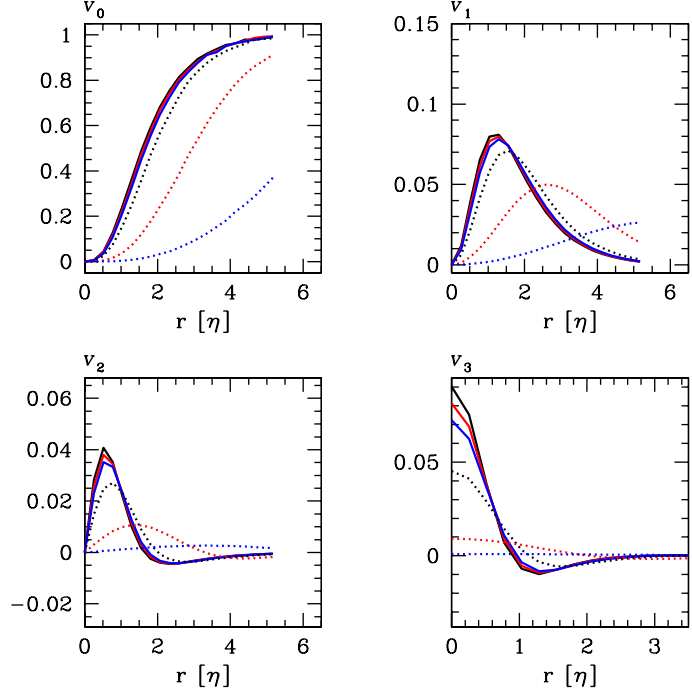}
  \caption[]{(color) Volume densities of the Minkowski functionals
    $v_{\mu}(r)$, $\mu = 0,1,2,3$ for the data set with $\beta=0$ and
    $St=0.6$. We compare the Minkowski functionals for samples with 
    a different number of points, obtained by randomly thinning the 
    original data set with 100k points.
    100k points (black solid), 90k points (red solid), 80k points (blue solid), 
    50k points (black dotted), 10k points (red dotted), 1k points (blue dotted). 
  }
\label{min-cmp-number}
\end{center}
\end{figure}

Minkowski functionals calculated from points decorated with spheres do
depend on the number density of the point distribution. One can derive
explicit expression in terms of high order correlation functions
quantifying the non-trivial dependence on the number density (see
e.g.~\cite{mec00}). In our analysis we always compare data sets with
the same number of points.  
However one may ask the question how well our Minkowski functionals
are converged.  In Fig.~\ref{min-cmp-number} the Minkowski functionals
for a randomly subsampled data set are shown.  Considering only 80\%
of the points we obtain nearly identical result as for the full
sample. This can be understood as follows: We decorate the particles
with spheres. In our case the particles follow well defined
structures.  For a radius equal to
zero, $v_3$ equals the number density.
That is why the $v_3$ curves fan out for small radii. Already for
$r\approx0.5\eta$ the curves for 100\%, 90\% and 80\% overlap again.
In this sense our results are well converged.

Obviously, the convergence must get lost
 if we further and further reduce the number of points.
Randomly subsampling (thinning) will inevitably lead to a convergence
towards Poisson (see the blue dotted line in fig.\ \ref{min-cmp-number})
-- the discreteness effects become more and more
important.

\begin{figure}
\begin{center}
\includegraphics[scale=0.5]{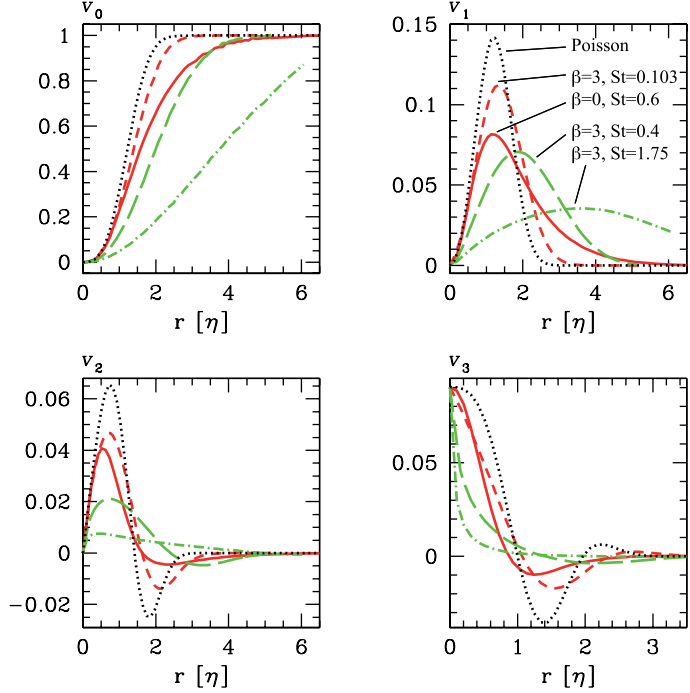}
\caption[]{(color) Volume densities of Minkowski functionals
$v_{\mu}(r)$, $\mu = 0,1,2,3$ for heavy particles with $\beta=0$ and
$St=0.6$ (red solid) and for light particles with $\beta=3$ and
$St=0.103$ (red short dashed): Whereas for these two particle types
$D_{KY}$ basically is the same, the Minkowski functionals clearly
differ.
As a second pair with nearly identical $D_{KY}$ but different
Minkowski functionals we compare two light particle distribution with
differing Stokes parameter: $\beta=3$, $St=0.4$ (green long dashed)
against $\beta=3$, $St=1.75$ (green dashed-dotted).
Poisson distribution behavior is reported for reference (black
dotted).}
\label{min-cmp}
\end{center}
\end{figure}

Up to now we investigated three extreme cases with $St=0.6$ and
$\beta=0,1,3$.  These three particle distributions also show different
$D_{KY}$, hence their morphological differences do not come as a
surprise. Now we investigate the morphology of the particle
distributions with similar $D_{KY}$ but quite different
physical parameters $\beta$ and $St$.
In figure~\ref{min-cmp} we compare heavy particles ($\beta=0$,
 $St=0.6$)
with $D_{KY}=2.59$ with light bubbles ($\beta=3$, $St=0.103$) with
$D_{KY}=2.55$.  Clearly, their morphological properties differ.
Although both have similar $D_{KY}$, the distribution of the heavy
particle shows a stronger deviation from the Poisson distributed
points than the light bubbles.
As another example, in figure~\ref{min-cmp} we also compare the
distribution of two sets of light bubbles, both with $\beta=3$ and
identical $D_{KY}=1.65$, but with different Stokes number. The data
set with $St=1.75$ shows a distribution dominated by isolated clusters
whereas the data set with $St=0.4$ allows for some interconnected
empty tunnels.

We now address the Reynolds number dependence of 
the results from the morphological analysis.
We compare the results 
for particles in turbulent flow with $Re_\lambda=75$
with simulations at higher Reynolds numbers $Re_\lambda=180$ and
$Re_\lambda=400$, keeping the parameters $\beta=0$ and $St=0.6$ fixed.
\begin{figure}
\begin{center}
\includegraphics[scale=0.5]{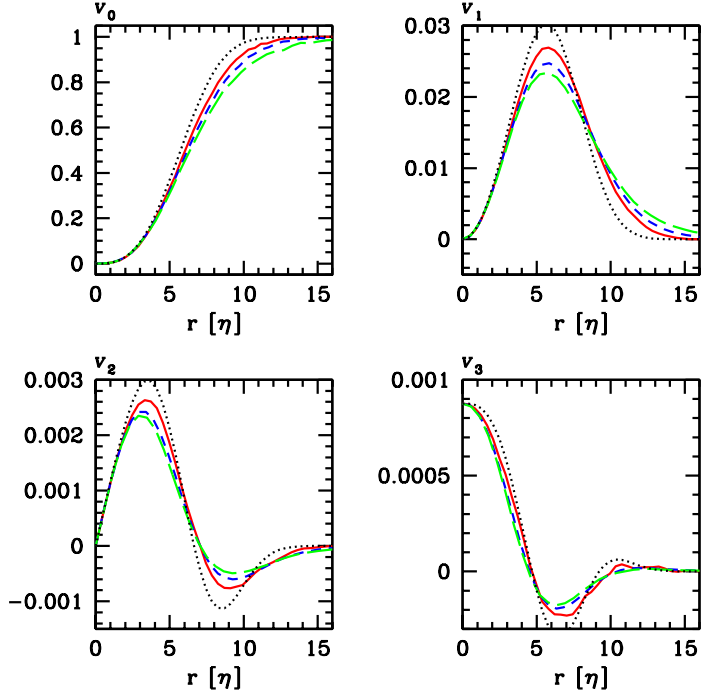}
\caption[]{(color) Volume densities of the Minkowski functionals
$v_{\mu}(r)$, $\mu = 0,1,2,3$ for particles with $\beta=0$ and
$St=0.6$ in simulations with different Reynold numbers:
$Re_\lambda=75$ (red solid), 
$Re_\lambda=180$ (blue short dashed),
$Re_\lambda=400$ (blue long dashed) .
Poisson distribution behavior is reported for reference (black
dotted). All point sets have been subsampled to the same number
density.}
\label{min-cmp-reynold}
\end{center}
\end{figure}
Only a weak dependence of the Minkowski functionals on the turbulence
intensity can be seen in Fig.~\ref{min-cmp-reynold}. A higher Reynolds
number consistently corresponds to a slightly more pronounced
deviation from the Poisson behavior. We observe similar results for
the bubble distribution with $\beta=3$, $St=0.6$.
This result agrees with the very weak Reynolds dependence of $D_{KY}$
already revealed 
in Fig.\ \ref{dkycut} and  in \cite{bec06b}.


\section{Conclusions and outlook}
Often the small-scale behavior in a turbulent dispersed multiphase
flow is of interest. Then scaling or contraction indices like the
Kaplan--Yorke Dimension $D_{KY}$ are the main tools. With $D_{KY}$ as
a function of $\beta$ and $St$ a simple picture emerges: For heavy
particles ($\beta<1$) we observe in the extreme cases a scaling
reminiscent of local planar structures, for light bubbles ($\beta>1$)
we find indications for local linear structures. The strength of the
deviation from $D_{KY}=3$ is further modulated by the Stokes parameter
$St$.
If one is interested in the connectivity and other (global)
morphological features of the particle distribution, the picture
becomes significantly more complex. The distribution of heavy
particles seems to allow interconnected empty tunnels, whereas the
distribution of bubbles typically shows isolated filamentary
structures. But, depending on both $\beta$ and $St$, also
interconnected structures appear in the bubble distribution. 
Using Minkowski functionals as morphological order parameters we are
able to quantify these geometrical and topological features in a
unique way.  Especially, we are able to overcome the degeneracy seen in
$D_{KY}$ as a function of $\beta$ and $St$.

The next step is to use these tools to compare data obtained by particle
tracking velocimetry (PTV) with those obtained from numerical simulations,
in order to test whether effective force models lead to the same particle
distribution as measured in experiment. 
To facilitate such an analysis, we make the code to calculate 
Minkowski functionals available to the general public, see
\url{http://www.mathematik.uni-muenchen.de/~kerscher/software/}.
Raw data can be downloaded 
 from the International CFD database, iCFDdatabase (\url{http://cfd.cineca.it}).

\vspace{0.3cm}
\noindent
{\it Acknowledgements:}
We thank Herbert Wagner and Massimo Cencini for interesting
and helpful discussions and the computer centers 
CASPUR (Rome), SARA (Amsterdam), and CINECA (Bologna) for CPU time.


\end{document}